\documentclass{article}

\usepackage{PRIMEarxiv}

\usepackage[utf8]{inputenc} 
\usepackage[T1]{fontenc}    
\usepackage{hyperref}       
\usepackage{url}            
\usepackage{booktabs}       
\usepackage{amsfonts}       
\usepackage{nicefrac}       
\usepackage{microtype}      
\usepackage{lipsum}
\usepackage{fancyhdr}       
\usepackage{graphicx}       
\graphicspath{{media/}}     

\title{How are the people in the photos judged? Analysis of brain activity when assessing levels of trust and attractiveness
}

\author{
  Bernadetta Bartosik, Grzegorz M. Wojcik, Andrzej Kawiak \\
  Department of Neuroinformatics and Biome \\
  Maria Curie-Sklodowska University \\
  Lublin, Poland\\
  \texttt{gmwojcik@live.umcs.edu.pl}
   \And
  Aneta Brzezicka \\
  Neurocognitive Research Center \\
  SWPS University of Social Sciences and Humanities \\
  Warsaw, Poland\\
}

\begin{document}
\maketitle

\begin{abstract}
Trust is the foundation of every area of life. Without it, it is difficult to build lasting relationships. Unfortunately, in recent years, trust has been severely damaged by the spread of fake news and disinformation, which has become a serious social problem. In addition to trust, the factor influencing interpersonal relationships is perceived attractiveness, which is currently created to a large extent by digital media. Understanding the principles of judging others can be helpful in fighting prejudice and rebuilding trust in society. One way to learn about people's choices is to record their brain activity as they make choices. The article presents an experiment in which the faces of different people were presented, and the participants' task was to assess how much they can trust a given person and how attractive they are. During the study, the EEG signal was recorded, which was used to build models of logistic regression classifiers. In addition, the most active areas of the brain that participate in the assessment of trust and attractiveness of the face were indicated.
\end{abstract}


\section{Introduction}
Since time immemorial, humans have been building first impressions about others based on appearance, and mainly on the face. In addition to basic information such as gender, age, emotions or ethnicity, more complex attributes such as intelligence, attractiveness, trustworthiness can also be read from the face (Koscinski, 2007; Oosterhof and Todorov, 2009). It is not always the first judgment that reflects the true face of the other person, but in many situations the first impression can be decisive. One of the primary factors swimming around in social relationships is the relationship between trustworthiness and beauty. Following the stereotype of " what is beautiful is good", it has been proven that attractive people are more likely to be considered trustworthy (Shinners, 2009). Another important factor is the emotions expressed. Individuals who exude positive energy and whose faces show joy will be judged as more trustworthy compared to those expressing sadness or anger at any given time (Sutherland et al., 2017). Additionally, gender is a factor that can affect trust. According to research, women and people with feminine or childlike facial features are given higher trust (Zebrowitz et al., 2015).\par

In everyday life, trust in another person plays a very important role during interpersonal interactions. The simplest example can be found in various social groups, in which people with trustworthy faces gain more approval (Tracy et al., 2020). Similar correlations can be observed with respect to people with attractive faces to whom it is easier to get along in a group and have greater support (Dion et al., 1972). The behavior of the business and financial world also correlates with the findings on trust. It has been shown that during trust games, participants are willing to invest more money if they play with a person who looks trustworthy (Van't Wout and Sanfey, 2008; Chang et al., 2010). Comparable observations apply to the online sales market. A seller who includes a trustworthy photo next to his listing is more likely to make a sale, even if he doesn't have reviews (Ert et al., 2016). Those who appear trustworthy are more likely to get a positive response against submitted credit applications (Duarte et al. (2012). Inferences drawn from facial appearance form the basis of social judgments. Offenders with untrustworthy faces experienced more severe punishments compared to those judged trustworthy (Wilson and Rule, 2015; Ancans and Austers, 2018). A similar relationship was indicated when evidence of the crime committed was insufficient (Porter et al., 2010). Attractive criminals have been shown to receive more lenient sentences compared to unattractive ones (Umukoro, Egwuonwu, 2014; Beaver at al., 2019). \par

The past few years have been quite a challenge for society. During the pandemic, all sorts of restrictions and obligations were imposed. One of them was the order to wear masks in public spaces. Partially covering the face makes it more difficult to read some of the visual information just read from the face, and thus can lead to feelings of insecurity (Hall et al., 2007), which in turn can contribute to reduced levels of trust (Acar-Burkay et al., 2014). The results of the study did not show a decrease in trust in people wearing a face mask (Grundmann et al., 2021). There is evidence indicating an increase in perceived trustworthiness towards strangers (Cartaud et al., 2020).\par

From a very young age, humans have the ability to recognize faces (De Heering, Rossion, Maurer, 2012; Jessen, Grossmann, 2019; Mondloch, Gerada, Proietti, Nelson, 2019), and the time needed to detect a face (depending on the situation) is just over 100 ms (Crouzet, Kirchner, Thorpe, 2010; Martin, Davis, Riesenhuber, Thorpe, 2018). It has been proven that the brain is capable of recognizing different features in stages. It first recognizes external appearance features, such as gender, and then analyzes complex features describing personality, attractiveness, or emotional state (di Oleggio Castello, Gobbini, 2015; Dobs, Isik, Pantazis, Kanwisher, 2019).\par

During face viewing, different areas of the brain responsible for processing other features are activated. In credibility assessment tasks, elevated neuronal activity is most often recorded from the amygdala, the ventromedial prefrontal cortex and the interior insula. The amygdala is considered one of the main areas associated with the analysis of emotional and social stimuli, as well as during credibility assessment (Costafreda et al., 2008). For credible-looking faces, amygdala activity decreases, while it increases for unreliable faces (Haas et al., 2015). Damage to the amygdala results in impaired assessment of face credibility (Adolphs et al., 1998).\par

The issue of trust and attractiveness discussed has an impact on relationships between people and on decision-making. The subject of the study is the brain activity during the evaluation of the level of trust and attractiveness of people depicted in pictures. This article presents an experiment to identify the areas of the brain that are most active in the process of assessing the level of trust and attractiveness.

\section{Design of the experiment}
During the study, participants' brain activity was recorded. The task of the participants in the study was to answer a question about trust and attraction towards the people in the photos. The purpose of the study conducted was to record brain activity.\par

\subsection{Photo database}
The experiment was prepared based on photos of the faces of men and women. Of the many sets, only those containing photos of real people were included. Artificially generated photo bases were omitted, as they can affect the level of reliability (Balas, Pacela, 2017). Photos were downloaded from online databases available for scientific research. The first is the Development Emotional Faces Stimulus Set (DEFSS) database (Meuwissen et al., 2017). The collection consists of 404 photos of men and women up to the age of 30. The photos show faces expressing various emotions and natural facial expressions. The photos were vetted for the emotions depicted before publication, which is an added benefit. The second dataset is the Multi - Racial Mega - Resolution (MR2) dataset (Strohminger et al., 2016). The MR2 database contains photos of the faces of 74 people who do not express emotions. The collection contains photos of people of different origins. A distinction can be made here between persons of European, Asian and African race. The photos have been verified by age gender and race, among other factors.\par
\begin{figure}[ht]
    \centering
    \includegraphics[scale=0.3]{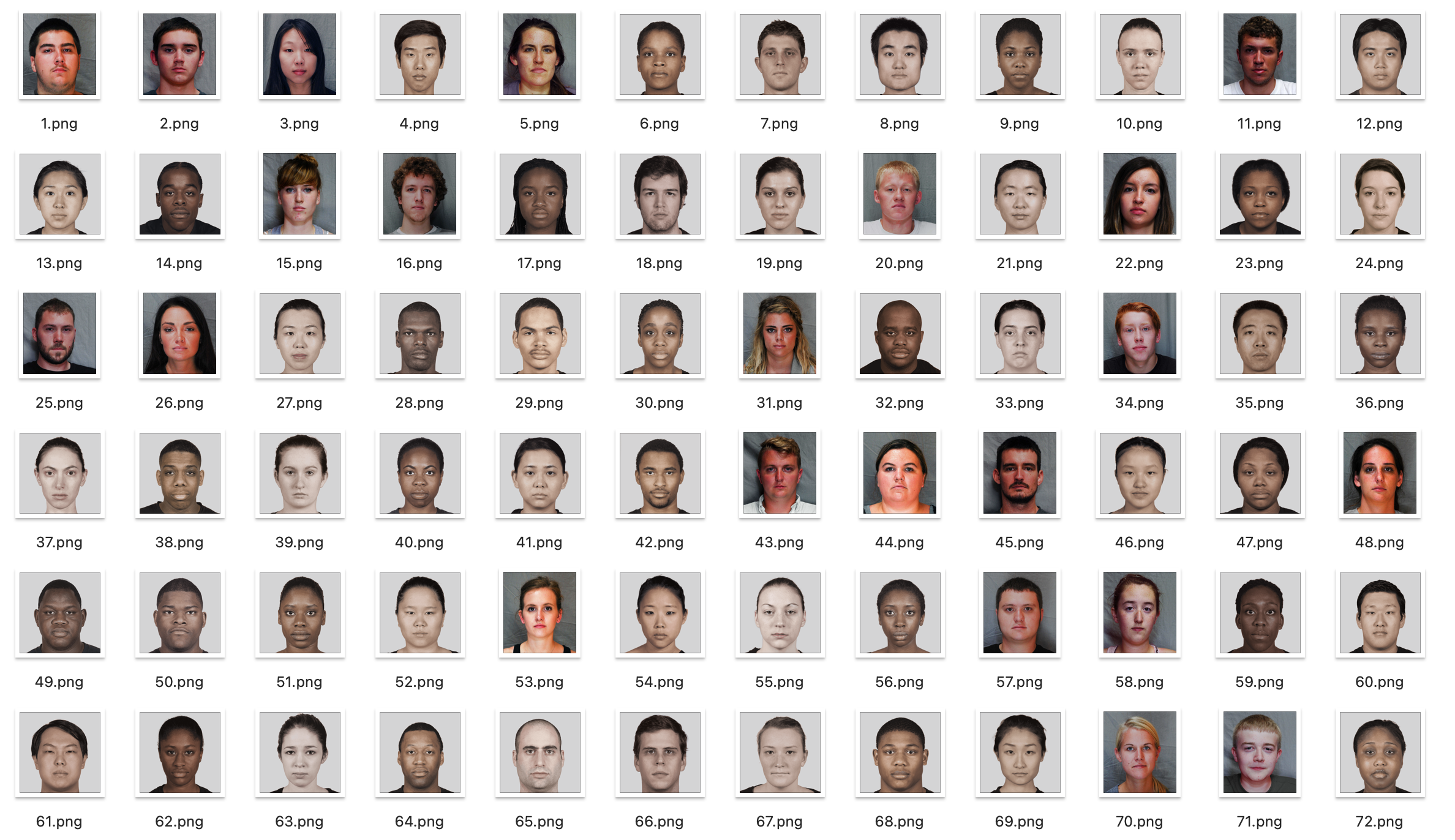}
    \caption{A collection of images going into the study.}
    \label{fig:my_label}
\end{figure}

For the purposes of the study, only photos showing neutral facial expressions that do not suggest any emotion were selected. One of the criteria for the selection of photos was to present the face from the front, so that the entire face is visible. Taking into account the indicated criteria, 100 photos of faces were selected. Maintaining gender balance, half of the photos present women's faces and another half present men's faces. Figure \ref{fig:my_label} shows some of the selected images for the experimental base.

\subsection{Pilot study}
During the pilot study (Bartosik et al, 2021), a survey was conducted in which participants rated the level of confidence and attractiveness of the people in the photos. The survey was conducted using the aforementioned facial photos. As a result of statistical analysis, four groups of photos were identified: attractive and trustworthy, attractive and untrustworthy, unattractive and trustworthy, unattractive and untrustworthy. Each listed group consists of six photos with the highest ratings among the photos, racial and gender division was taken into account.

\subsection{Experiment}
\begin{figure}[ht]
    \centering
    \includegraphics[scale=0.8]{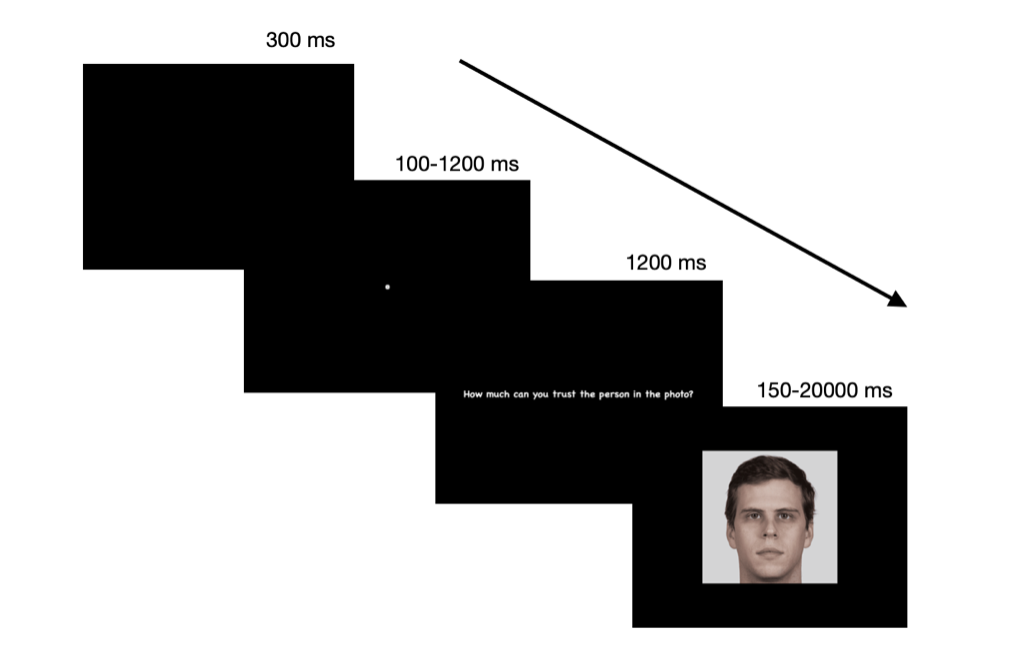}
    \caption{The course of the experiment.}
    \label{fig:my_label1}
\end{figure}

The task of the participants in the experiment was to indicate the level of confidence and attractiveness of the person depicted in the photo. To design the experiment, 26 photos were selected, which were extracted based on statistical analysis in a pilot study. The design work was carried out using OpenSesame software. Figure \ref{fig:my_label1} shows the scheme of the experiment. The participant in the study went through the stages one by one starting with a black screen, a fixation point, a question card and a card with the photo to be evaluated. The display time of the fixation point and the photo varied randomly between 100 - 1200 ms for the fixation point and 150 - 20000 ms for the photo. The study participant answered two questions: "How much are you able to trust the person in the photo?" and "How attractive is the person in the photo?" using a five-point Likert scale, where a value of five means most and a value and one means very little. The participant in the study answered the questions for each picture presented.

\subsection{Participants in the experiment}
The study involved 61 young people, students of computer science at the Maria Curie-Skłodowska University in Lublin. People aged 18-23 were invited. Registration for the study took place via an online form, in which head circumference and time availability had to be provided. To ensure the confidentiality of personal information, each participant was assigned a randomly generated ID number. In order to create the most representative research group, students were recruited according to specific criteria. It was assumed that right-handed people with short hair could participate in the study, because long hair causes more noise in the signal. Due to the low number of short-haired women in computer science, only men were recruited for the study. Only people without permanent or serious health problems within a year that would hinder the conduct of the study or affect the quality of the collected data could participate in the study. As an element of preparation for the study, participants were asked not to consume alcohol for at least three days before the planned study.
\clearpage

\subsection{EEG signal recording}
High-quality equipment distributed by Electrical Geodesic System Inc (EGI) was used to record the EEG signal. The measurement laboratory was equipped with an amplifier that allows signal recording through 256 channels (HydroCel GSN 130 Geodesic Sensor Net) at a frequency of up to 500 Hz. Signal preprocessing including removal of artifacts such as eye blinks and eye movements was carried out using EGI system software scripts. The prepared EEG signal was subjected to further calculations. During the experiment, the signal from 256 electrodes was collected. The ERP signal was determined for each electrode. Based on the determined ERP signal and source localization techniques, the mean electrical charge (MEC) that flows through the BA placed under the electrodes during the CPTI cognitive processing time interval was estimated. Using the determined ERP signals, a sLORETA source localization analysis was performed.

\section{Results of the experiment}
\subsection{Statistical method}
Analysis of the collected data was carried out using a logistic regression model, which easily describes the relationship between the dependent variables in dichotomous form and the explanatory variables. Two issues were used in the experiment: trust and attractiveness, so two sets of data were obtained. Using the selected model, an analysis was performed for both sets. In the first set, the dependent variable is whether the research participant trusted or distrusted and the descriptor variables are the different brain areas. In the second set of data, the describing variables remain the same, the dependent variable changes, which takes two states: attractive, not attractive.\par

The logistic regression model was built in Python (version 3.9.9) using the scikit-learn library (version 1.2.2)

\subsection{Data preparation}
Before classifying the data, data sets had to be prepared. The first step was the selection of the time interval. The average activity of brain areas over time was determined. The Desikan-Killiany atlas was used in the sets. Based on the time course of the signal, the sections with higher brain activity were selected. In both data sets, the highest brain activity was observed in the intervals between 250 and 350 ms after the stimulus onset. The second data preparation step was to determine the average electrical charge with respect to the obtained time intervals. The final set contained the MEC values in the time interval for each brain area. The next step was to divide the data into learning and test sets. Sixty subjects participated in the experiment. For each person, two events were determined for each dependent variable, that is, for the trust variable: trusted, distrusted, and for the attractiveness variable: attractive, unattractive. This yielded about 120 results for each dependent variable (in the case of the trust variable, several events were omitted due to insufficient quality). A standard 80:20 split was used for data selection, where 80\% was the data of the learning set, 20\% of the validation set. The final step was to reduce the number of independent variables so as to pinpoint the most important decision-making areas to the brain. The Desikan-Killiany atlas that was used divides the brain into 34 regions, while each region is divided into two hemispheres, so that the data initially contained 68 descriptor variables. Optimization of the number of variables was carried out using the Recursive Feature Elimination with Cross-Validation (RFECV) method. This is an algorithm that allows the least significant features to be eliminated recursively using cross-validation to select the best subset.

\subsection{Data classification}

\begin{figure}[ht]
    \centering
    \includegraphics[scale=0.8]{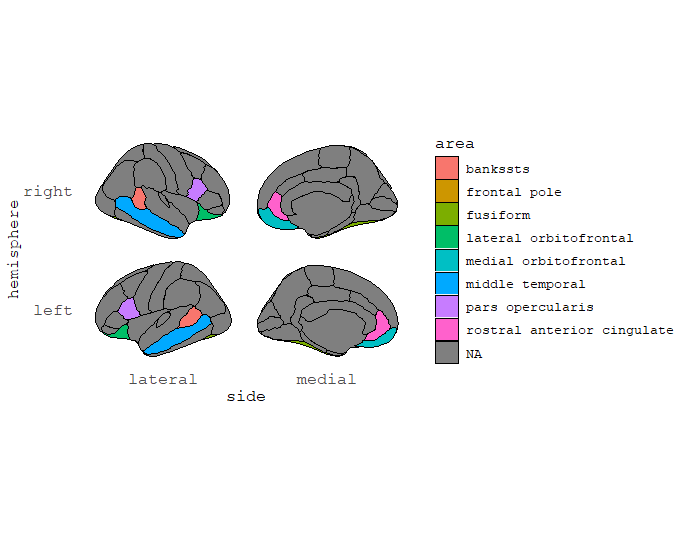}
    \caption{The most significant brain areas for the "Trust" model.}
    \label{fig:my_label6}
\end{figure}

In the process of eliminating the number of descriptive variables, it was possible to identify the more relevant brain areas influencing the participant's decision. In the case of the model based on the dependent variable "confidence", from the original 68 detailed areas detailing the right and left hemispheres (34 total areas), it was possible to reduce the number of variables to 10 detailed areas: bankssts L, bankssts R, frontalpole L, fusiform R, lateral orbitofrontal R, medial orbitofrontal L, medial orbitofrontal R, middle temporal L, pars opercularis R, rostral anterior cingulate L (8 general areas: bankssts, frontalpole, fusiform, lateral orbitofrontal, medial orbitofrontal, middle temporal, pars opercularis, rostral anterior cingulate) (Figure \ref{fig:my_label6}). \par
\begin{figure}[ht]
    \centering
    \includegraphics[scale=0.6]{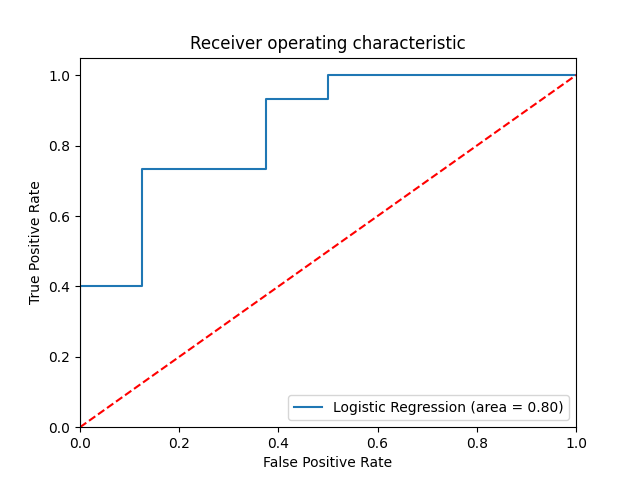}
    \caption{ROC curve for the "Trust" model.}
    \label{fig:my_label2}
\end{figure}

\begin{figure}[ht]
    \centering
    \includegraphics[scale=0.6]{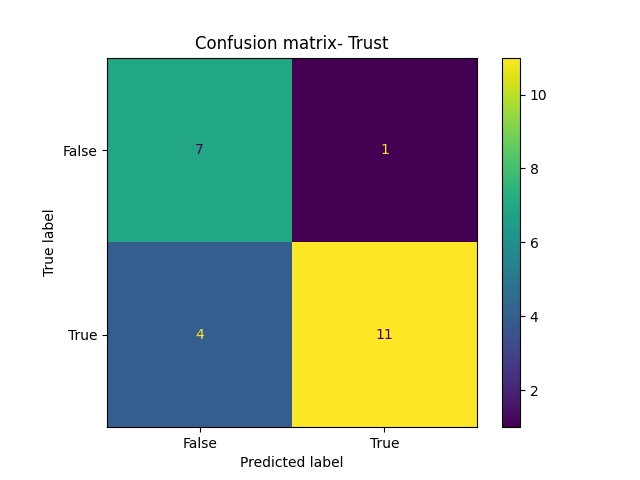}
    \caption{Confusion matrix for the "Trust" model.}
    \label{fig:my_label3}
\end{figure}

Based on these 10 areas, a classifier was built that makes it possible to predict the participant's decision with satisfactory efficiency. Accuracy for the obtained model is 0.78. Below (Figure \ref{fig:my_label2}, \ref{fig:my_label3}) are graphs describing the model of the "trust" variable. The classification quality measures presented in the table \ref{tabAcc} confirm that a good quality classifier was successfully built to predict trust ratings based on personality traits.\par
\begin{figure}[ht]
    \centering
    \includegraphics[scale=0.8]{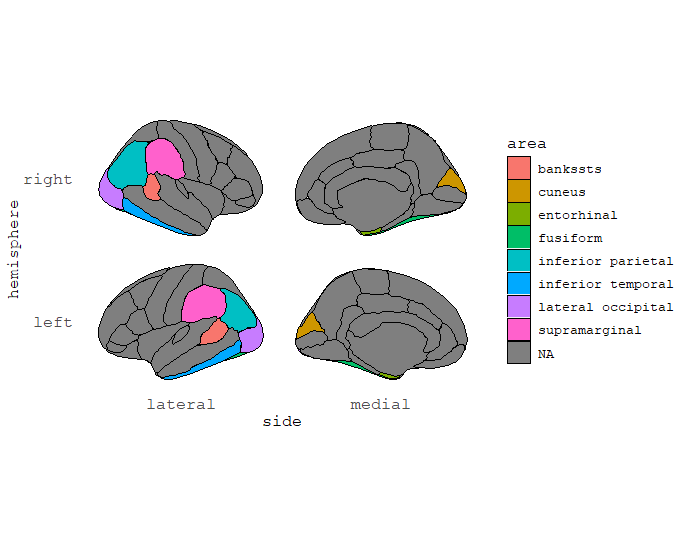}
    \caption{The most significant brain areas for the "Attractiveness" model.}
    \label{fig:my_label7}
\end{figure}

\begin{table}[!ht]
 \centering
\begin{tabular}{|c|c|c|c|}
    \hline
    Accuracy & Precision & Recall & F1-score \\ \hline
    0.78 & 0.92 & 0.73 & 0.81 \\ \hline
\end{tabular}
\caption{Classification measures that represent classifier quality for trust}
\label{tabAcc}
\end{table}
\clearpage
During the elimination of features for the model based on the dependent variable "attractiveness" it was possible to extract 8 areas of detail: bankssts R, cuneus R, entorhinal L, fusiform L, inferior parietal R, inferiort emporal L, lateral occipital L, supramarginal R (8 general areas: bankssts, cuneus, entorhinal, fusiform, inferior parietal, inferiort emporal, lateral occipital, supramarginal)(Figure \ref{fig:my_label7}). \par

\begin{figure}[ht]
    \centering
    \includegraphics[scale=0.6]{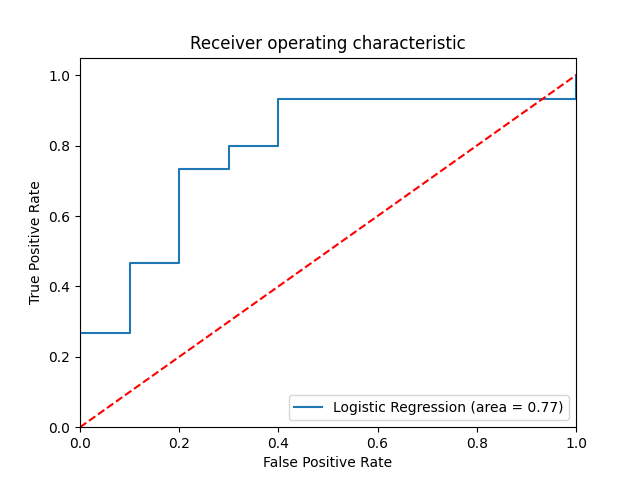}
    \caption{ROC curve for the "Attractiveness" model.}
    \label{fig:my_label4}
\end{figure}

\begin{figure}[ht]
    \centering
    \includegraphics[scale=0.6]{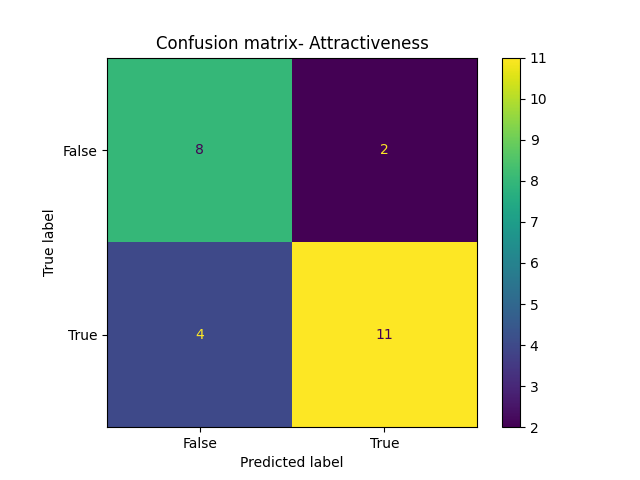}
    \caption{Confusion matrix for the "Attractiveness" model.}
    \label{fig:my_label5}
\end{figure}

The classifier, which was built on the basis of the most relevant descriptive variables, obtained Accuracy of 0.76. Figure \ref{fig:my_label4}, \ref{fig:my_label5} shows the characteristics of the built model. The classification quality measures presented in the table \ref{tabAccAtr} confirm that a good quality classifier was successfully built to predict attractiveness ratings based on personality traits.\par 
\begin{table}[!ht]
 \centering
\begin{tabular}{|c|c|c|c|}
    \hline
    Accuracy & Precision & Recall & F1-score \\ \hline
    0.78 & 0.92 & 0.73 & 0.81 \\ \hline
\end{tabular}
\caption{Classification measures that represent classifier quality for attractiveness}
\label{tabAccAtr}
\end{table}
\clearpage

In the process of data preparation, the most significant descriptor variables were determined for each dependent variable. Based on these variables, it can be determined which brain regions are most active during each decision. The bankssts and fusiform regions appear in both models, indicating that for both trust and attraction decisions, these are active areas.

\section{Discussion}
The most relevant brain regions that are involved in decision-making were verified. The study included the collection of information on trust and attraction, qualities that come into play on a daily basis and often determine the success of a continuing relationship. Stimulus stimulation activates different areas of the brain to process the information. In the experiment, participants were stimulated by displaying pictures of men and women. The participant's task was to indicate the level of confidence in the depicted face and determine whether the face was attractive or not. The recorded brain activity was used to build classifiers predicting confidence ratings and attractiveness ratings. The feature elimination algorithm listed eight brain regions each for the two dependent variables of greatest importance during stimulus analysis. The bankssts, frontalpole, fusiform, lateral orbitofrontal, medial orbitofrontal, middle temporal, pars opercularis, and rostral anterior cingulate regions have the greatest influence on the trusted/un-trusted classification. The attractive/unattractive classification was based on the bankssts, cuneus, entorhinal, fusiform, inferior parietal, inferiort emporal, lateral occipital, supramarginal regions. Most of these regions are responsible for processing social signals, processing faces, evaluating emotions.\par
Bankssts is an area encompassing the superior temporal sulcus (STS), which is responsible for, among other things, processing social signals such as processing faces, credibility, intentions (Ethofer et al., 2006). The fusiform area, also known as the fusiform gyrus, is a region of the brain located in the temporal lobe, near the occipital lobe. It is primarily responsible for visual processing and recognition of faces and other complex visual stimuli such as objects, animals, and words (Rangarajan et al., 2014). The occipital area, also known as the occipital lobe, is a region of the brain located at the back of the head, behind the parietal and temporal lobes. It is primarily responsible for visual processing and perception, including the interpretation of color, shape, movement, and depth (Nagy et al., 2012). The first two areas are among the variables describing the "trust" model, while the last two are among those variables of the "attraction" model. All areas are involved in the processing of visual stimuli. According to Haxby et al. the STS, OFA (occipital face area) and FFA (fusiform face area) areas, which are part of the above-described regions, constitute the perceptual face processing system.\par
The frontal pole, also known as the rostral prefrontal cortex, is a region of the brain located in the front of the frontal lobe, at the very top of the brain. It is believed to play a key role in executive functions, such as planning, decision-making, working memory. Damage to the vmPFC area, which is included in this region, results in deficits in social function (Moretti et al., 2009). The middle temporal area, also known as the middle temporal gyrus, is a region of the brain located in the temporal lobe, just above the fusiform gyrus. It is primarily responsible for visual motion processing and object recognition. The inferior temporal region, also known as the inferior temporal lobe, is an area of the brain located in the temporal lobe, below the medial temporal lobe. It is primarily responsible for high-level visual processing, including object and face recognition (Perrett et al., 1982), categorization and visual memory. The medial orbitofrontal cortex, also known as the medial prefrontal cortex, is a region of the brain located in the frontal lobe, just above the eyes. It is involved in a variety of cognitive and emotional processes, including decision-making, reward processing, social behavior, and emotion regulation. The rostral anterior cingulate cortex, also known as the dorsal anterior cingulate cortex, is a region of the brain located in the frontal lobe, just behind the medial prefrontal cortex. It is involved in a variety of cognitive and emotional processes, including attentional control, conflict monitoring, decision-making, and emotion regulation. The supramarginal gyrus is a part of the parietal lobe of the brain located in the posterior portion of the lateral sulcus, also known as the Sylvian fissure. It participates in social cognition, including emotion recognition and empathy, but this is not its main function. The cuneus is a brain region located in the occipital lobe, which is situated at the back of the brain. The cuneus plays an important role in visual processing, particularly in the processing of visual information from the eyes. It is involved in the early stages of visual processing, such as the recognition of basic features like lines and edges, and the detection of visual motion.
\section{Conclusions}
The above study documents that the use of source location algorithms (sLORETA) and machine learning classifiers make it possible to predict trust ratings and attractiveness ratings with fairly high accuracy. It was verified which areas are significant depending on the dependent variable. A previous study (Bartosik et al, 2021) focused on presenting the most important personality traits in the process of trust and attractiveness ratings. In the future, it is planned to test whether and which personality traits influence decision-making based on brain activity.


\end{document}